\documentclass[a4paper,11pt]{article}
\usepackage{amsfonts}
\usepackage[T1]{fontenc}
\usepackage{graphicx}
\usepackage[colorlinks]{hyperref}
\usepackage{amsmath}
\usepackage[psamsfonts]{amssymb}
\usepackage{epsf,epsfig}
\usepackage{amsmath }
\usepackage{amssymb}
\usepackage{latexsym}
\usepackage{color}
\usepackage{cite}
\usepackage{float}
\usepackage{epsf,epsfig}
\usepackage[psamsfonts]{amssymb}
\usepackage{hyperref}
\usepackage[textfont={footnotesize,sf},labelfont={color=blue,bf,sf},labelsep=endash]{caption}
\usepackage[position=top,labelfont={color=blue,bf,sf}]{subfig}
\usepackage{bm}
\usepackage{makeidx}
\usepackage{colortbl}
\usepackage{csquotes}
\usepackage[squaren]{SIunits}
\newcommand{\bra}{\begin{array}}
\newcommand{\era}{\end{array}}
\newcommand{\beq}{\begin{equation}}
\newcommand{\eeq}{\end{equation}}
\newcommand{\beqar}{\begin{eqnarray}}
\newcommand{\eeqar}{\end{eqnarray}}

\def\BC{\bb C}
\def\_\BC{\bbi C}

\def\Tr {{\rm Tr}}

\def\( {\left(}
   \def\) {\right)}
\def\[ {\left[}
\def\] {\right]}

\def\Tr {{\rm Tr}}

\newcommand{\lb}{\label}

\def\eqref#1{\textcolor{blue}{(\ref{#1})}}
\usepackage{amsmath}
\usepackage{xcolor}
\colorlet{linkequation}{blue}
\usepackage[colorlinks]{hyperref}

\newcommand*{\SavedEqref}{}
\let\SavedEqref\eqref
\renewcommand*{\eqref}[1]{%
  \begingroup
    \hypersetup{
      linkcolor=linkequation,
      linkbordercolor=linkequation,
    }%
    \SavedEqref{#1}%
  \endgroup
}
\hypersetup{linkcolor=black}
\begin{document}
\begin{titlepage}
\setcounter{page}{1}
\renewcommand{\thefootnote}{\fnsymbol{footnote}}


\vspace{5mm}
\begin{center}

{\Large \bf {Quantum teleportation via a two-qubit Heisenberg  XXX chain with x-component of Dzyaloshinskii-Moriya  interaction }}

\vspace{4mm}

{\bf Rachid Hou\c{c}a$^{1,2}$\footnote{r.houca@uiz.ac.ma}
,
Abdelhadi Belouad$^{2}$\footnote{belabdelhadi@gmail.com}
,
El Bouâzzaoui Choubabi$^{2}$\footnote{choubabi.e@ucd.ac.ma}
,
Abdellatif Kamal$^{2,3}$\footnote{abdellatif.kamal@ensam-casa.ma}
and
Mohammed El Bouziani$^{2}$\footnote{elbouziani.m@ucd.ac.ma}
}

\vspace{3mm}

{$^{1}$\em LPTHE Laboratory, Department of Physics, Faculty of Sciences, Ibn Zohr University, PO Box 8106, Agadir, Morocco}

{$^{2}$\em Team of Theoretical Physics, Laboratory LPMC, Department of Physics, Faculty of Sciences, Chouaib Doukkali University, PO Box 20, 24000 El Jadida, Morocco}

{$^{3}$\em ISPS2I Laboratory, National Higher School of Arts and Crafts (ENSAM), Hassan II University, 20670, Casablanca, Morocco}

\vspace{2.5cm}

\begin{abstract}
In this paper, we investigate the thermal entanglement and teleportation of a thermally mixed entangled via a two-qubit Heisenberg XXX chain with the x-component $D_x$ of the Dzyaloshinskii-Moriya interaction. Our findings suggest that temperature $T$,  spin coupling constant J, and the x-components $D_x$ may influence entanglement of the output states and, consequently, the possibilities of teleportation protocols. Furthermore, these results indicate that the entanglement of the output states requires a low-temperature regime, weak  Dzyaloshinskii-Moriya interaction, or an antiferromagnetic chain. Finally, the channel becomes entangled, making the teleportation protocol conceivable and feasible.
\end{abstract}
\end{center}

\vspace{2cm}



\end{titlepage}

\section{Introduction}
The phenomenon of quantum teleportation is, as the name suggests, utterly fascinating. Teleportation \cite{Ben1} is a striking manifestation of what is commonly referred to as the non-locality of quantum mechanics. It allows quantum information to be transmitted directly from port A to port B without any intermediate point. From a technological point of view, it is essential for everything related to quantum communications. One can imagine it at the heart of several quantum cryptography protocols and as a necessary component of the future quantum computer. Quantum cryptography makes it possible, among other things, to establish, in theory and conditionally to the validity of quantum mechanics, communication channels (classical or quantum) whose eavesdropping is impossible \cite{Ben2}.
On the other hand, the quantum computer can solve some unsolvable problems in a reasonable time for the classical computer, the one we use today. Shor's algorithm for factoring large prime numbers is a powerful example \cite{Shor}. Still, the quantum computer should go beyond that: if we are to believe Church-Turing-Deutsch's thesis, the quantum computer would effectively simulate any physical phenomenon\cite{loyd,Nie}. This fundamental task is currently impossible for the classical computer since affecting a quantum system requires classical exponential resources to the best of current knowledge.

It is essential to mention that while quantum mechanics is very unsettling at first glance, it is a physical theory with an excellent experimental basis. For example, Bell's inequalities \cite{bell}, which demonstrate the entanglement phenomenon, have been verified and rechecked repeatedly \cite{Asp3,Hensen,Tittel}, perhaps precisely because of their counterintuitive character. In recent years, quantum teleportation has also fallen out of the ranks of theoretical curiosities: it was carried out, to give only a few examples, between photons over distances of $600\;m$ below the Danube \cite{Urs} and $143\;km$ in the Canary Islands \cite{Ma}, as well as over a distance of $3\;m$ between sets of nitrogen-vacancy centers of two diamonds \cite{Pfaffb}.

A point common to all physical implementations of quantum teleportation is the fatality of noise which means that any concrete realization of theoretical protocols necessarily suffers from imperfections. In quantum computing, this difficulty is compounded by the fact that one cannot clone quantum information, which is very fragile and degrades during any interaction with its environment. The problem is that before the invention of quantum corrective codes \cite{Sho}, some scientists doubted that the quantum computer could ever be built. Hope has returned, but as evidenced by the current lack of a quantum computer, the fight against noise has yet to be won, and much remains to be learned.

Quantum entanglement is a phenomenon observed in quantum mechanics in which the quantum state of two systems must be described globally without being able to separate one system from the other, although they may be spatially separated. When two systems – or more – are placed in an entangled state, there are correlations between the observed physical properties of the two systems, which would not be present if one could attribute individual properties to each of the two systems $\xi_1$ and $\xi_2$. Consequently, even though large spatial distances separate them, the two systems are not independent, and we must consider $\{\xi_1+\xi_2\}$ as a single system. Wootters has shown that, in the case of a two-qubit system, the entanglement of a formation may be produced directly by applying the concurrence formula of the state to the system \cite{6,7}. Measurement-induced non-locality and Bell non-locality are investigated in a system of two coupled quantum dots entangled through their interaction with a cavity mode, including Förster and exciton–phonon interactions \cite{MOH1}. Also, the non-local correlations via Wigner-Yanase skew information in two superconducting-qubit having mutual interaction under phase decoherence are studied by A-B. A. Mohamed \cite{MOH2}. Recently in Ref. \cite{MOH3} the non-classical correlations are investigated for bipartite partitions of two qubits trapped in two spatially separated cavities connected by an optical fiber. Additionally, A-B. A. Mohamed et \textit{al}. have explored the quantum correlations for two coupled quantum wells. Each quantum well is inside a semiconductor microcavity connected by an optical fiber\cite{MOH4}.

The spin chain is one of the most obvious choices for the
realization of entanglement. The Heisenberg model is the most straightforward way of examining and
researching the behavior of spin chains. Heisenberg XXX-chain entanglement was explored for the
first time by Nielsen \cite{8}. He demonstrated that entanglement in
such systems happens only in the antiferromagnetic situation when the temperature is below a certain
threshold. Several studies have been conducted after Nielsen's work on entanglement in two-qubit systems such as the two-qubit XY, XXX, and XXZ systems in the existence of both a homogeneous and an inhomogeneous magnetic field \cite{9,10,11,12,13,14}. Anisotropy owing to spin coupling has also been investigated in the $x$, $y$, and $z$ directions, as has been done in several previous studies \cite{15, 16,houca}. For XYZ Heisenberg systems, Yang et al. Showed in ref \cite{17} that an inhomogeneous external magnetic field may improve the critical temperature and increase the entanglement. The spin-orbit (SO) interaction causes another sort of anisotropy \cite{18, 19, 20, 21, 22, 23, 24}. The influence of SO interaction on thermal entanglement in an XXZ system with two-qubit, without magnetic field, was published in \cite{25}.

According to C. Bennet et al. \cite{Ben1}, teleportation may be accomplished by combining two entangled but geographically distant particles. Theoretically, they claimed, less entangled states may still be used for teleportation. Afterward, S. Popescu demonstrated, using the hidden variable concept, that teleportation of a quantum state through pure classical transmission cannot be accomplished with the fidelity of more than $2/3$ \cite{Popescu}. To transport quantum information with the fidelity of more than $2/3$, mixed quantum channels with fidelity greater than $2/3$ are beneficial. When the channel is quantum mechanically correlated, J. Lee and M.S. Kim show that quantum teleportation preserves the character of quantum correlation in the unidentified entangled state. The following study entanglement teleportation using two copies of Werner states \cite{42}.

Given that the integrable models of magnetism in low-dimensional materials are based on the homogeneous Heisenberg XXX model \cite{Mota,houca}, our objective in this paper is to investigate and conclude results for the low-dimensional magnetism physics, which recently focused on finding novel magnetic compounds and refining their properties to fulfill emerging technology demands. To do this, we will consider a two-qubit Heisenberg XXX chain with an x-component Dzyaloshinskii-Moriya (DM) interaction to investigate the thermal entanglement teleportation. We reduce our Hamiltonian to identify the expressions of output concurrence and fidelity of teleportation protocol. To accomplish this, we diagonalize our system to discover the solutions to the energy spectrum, which lets us find the density matrix essential to identify the output concurrence and fidelity expressions.

The present paper is organized as follows. In section {\color{blue}2}, We quickly present the model Hamiltonian and the characteristics of the ground and thermal states. Section {\color{blue}3} will be devoted to the thermal entanglement teleportation by determining the concurrence of the output state and the fidelity. In section {\color{blue}4}, numerical studies will be performed to highlight the system behavior. Finally, the paper is closed with a findings overview and our perspectives.
\section{Theoretical model}
We consider a two-qubit Heisenberg XXX chain with the x-component of DM anisotropic antisymmetric interaction. The Hamiltonian of the system is given by
\beq\lb{fr}
\mathcal{H}=J\sum_n^N \left({\color{black}\sigma}_n^x{\color{black}\sigma}_{n+1}^x+{\color{black}\sigma}_n^y{\color{black}\sigma}_{n+1}^y+{\color{black}\sigma}_n^z{\color{black}\sigma}_{n+1}^z\right)+D_x \left({\color{black}\sigma}_n^y{\color{black}\sigma}_{n+1}^z-{\color{black}\sigma}_n^z{\color{black}\sigma}_{n+1}^y\right)
\eeq
$\sigma^{x,y,z}$ are the typical Pauli matrices, and $J$  represents the real coupling constant for the spin interaction. Antiferromagnetic chains are when the value of $J > 0$, while ferromagnetic chains are those in which the value of $J < 0$. The symbol $ D_x$ denotes the x-component of the DM interaction. Using the conventional computing base $|00>$, $|01 >$, $|10 >$, $|11 >$, the Hamiltonian equation \eqref{fr} for two-qubit can be expressed using its matrix form by
\beq\lb{1}
\mathcal{H}=\left(
\begin{array}{cccc}
 J & i D_x &  -i D_x & 0 \\
 -i D_x & -J & 2 J & i D_x \\
  i D_x & 2 J & -J & -i D_x \\
 0 &  -i D_x & i D_x & J \\
\end{array}
\right)
\eeq
The solution of the eigenvalue equation generates the eigenvalues listed below
\begin{eqnarray}\lb{22}
\epsilon_{1,2} &=&J \\ \nonumber
\epsilon_{3,4} &=& -J\pm 2 \sqrt{D_x^2+J^2}
\end{eqnarray}
as well as the eigenvectors that are connected to them
\begin{eqnarray}\lb{psi}
  |\varphi_1\rangle &=&  \frac{1}{\sqrt{2}}|00\rangle +\frac{1}{\sqrt{2}}|11\rangle \\ \nonumber
  |\varphi_2\rangle &=& \frac{1}{\sqrt{2}}|01\rangle+\frac{1}{\sqrt{2}}|10\rangle \\ \nonumber
  |\varphi_3\rangle &=& -\frac{1}{\sqrt{2}}\sin (\theta_1)|00\rangle{\color{red}+}\frac{i}{\sqrt{2}}\cos (\theta_1)|01\rangle{\color{red}-}\frac{i}{\sqrt{2}}\cos (\theta_1)|10\rangle+ \frac{1}{\sqrt{2}}\sin (\theta_1)|11\rangle \\ \nonumber
  |\varphi_4\rangle &=&-\frac{1}{\sqrt{2}}\sin (\theta_2)|00\rangle{\color{red}-}\frac{i}{\sqrt{2}}\cos (\theta_2)|01\rangle{\color{red}+}\frac{i}{\sqrt{2}}\cos (\theta_2)|10\rangle+ \frac{1}{\sqrt{2}}\sin (\theta_2)|11\rangle
\end{eqnarray}
where $\theta_{1,2}$ are determined by the formula
\begin{eqnarray}
  \theta_{1,2} &=&\arctan\left(\frac{D_x}{\sqrt{D_x^2+J^2}\mp J}\right)
\end{eqnarray}
After establishing the spectrum of our system, getting the density matrix is a straightforward procedure. The density matrix $\rho(T)$ may represent the state of a system at a given temperature $T$ when it is in thermal equilibrium. Indeed, the expression for $\rho(T)$ is as follows:
\beq
\rho(T)={1\over\mathbb{Z}}e^{-\beta \mathcal{H}}
\eeq
where
\beq
\mathbb{Z}=\Tr e^{-\beta \mathcal{H}}
\eeq
Where $\mathbb{Z}$ denotes the canonical ensemble partition function and $\beta={1/k_BT}$ indicates the inverse thermodynamic temperature, where $k_B$ represents the Boltzmann's constant, which is taken as unity in the following for the sake of simplicity. That will  be succeeded by using the spectral decomposition of the Hamiltonian \eqref{1}, which enables the thermal density matrix $\rho(T)$ to be expressed as
\beq\lb{3}
\rho(T)={1\over\mathbb{Z}}\sum_{l=1}^{4}e^{-\beta \epsilon_l}|\phi_l\rangle\langle\phi_l|
\eeq
It is possible to describe the density matrix of the system in the standard computational basis, as stated before in thermal equilibrium, by putting the equations \eqref{22} and \eqref{psi} into the equation \eqref{3} and getting the result
\beq\lb{43}
\rho(T)={1\over\mathbb{Z}}\left(
\begin{array}{cccc}
 a & i \mu & i \nu & c \\
 -i\mu & b & d & -i\nu \\
 -i\nu & d & b & -i\mu \\
 c & i\nu & i\mu & a \\
\end{array}
\right)
\eeq
the elements matrix is represented by the equations
\begin{eqnarray}\lb{23}
a &=&\frac{e^{-\beta \epsilon_1}}{2}+\frac{1}{2} e^{-\beta  \epsilon_3} \sin ^2(\theta_1)+\frac{1}{2} e^{-\beta  \epsilon_4} \sin ^2(\theta_2) \\ \nonumber
b &=& \frac{e^{-\beta  \epsilon_2}}{2}+\frac{1}{2} e^{-\beta \epsilon_3} \cos ^2(\theta_1)+\frac{1}{2} e^{-\beta \epsilon_4} \cos ^2(\theta_2) \\ \nonumber
c &=&\frac{e^{-\beta  \epsilon_1}}{2}-\frac{1}{2} e^{-\beta  \epsilon_3} \sin ^2(\theta_1)-\frac{1}{2} e^{-\beta  \epsilon_4} \sin ^2(\theta_2) \\ \nonumber
d &=&\frac{e^{-\beta  \epsilon_2}}{2}-\frac{1}{2} e^{-\beta  \epsilon_3} \cos ^2(\theta_1)-\frac{1}{2} e^{-\beta  \epsilon_4} \cos ^2(\theta_2)\\ \nonumber
\mu &=& \frac{1}{2} e^{-\beta  \epsilon_3} \sin (\theta_1) \cos (\theta_1)-\frac{1}{2} e^{-\beta  \epsilon_4} \sin (\theta_2) \cos (\theta_2) \\ \nonumber
\nu &=& \frac{1}{2} e^{-\beta  \epsilon_4} \sin (\theta_2) \cos (\theta_2)-\frac{1}{2} e^{-\beta \epsilon_3} \sin (\theta_1) \cos (\theta_1)
\end{eqnarray}
Consequently, the partition function is clearly described by
\beq
\mathbb{Z}=2 e^{-\beta  J}+2 e^{\beta  J} \cosh \left(2 \beta  \sqrt{D_x^2+J^2}\right)
\eeq
After determining all previously required ingredients, the next part will examine thermal entanglement teleportation by analyzing output state concurrence and the fidelity of entanglement teleportation.
\section{Thermal Entanglement teleportation}
\subsection{Concurrence of output state}
In the Heisenberg spin chain, it is feasible to conceive the thermal mixed state as a generalized depolarizing channel for the entanglement teleportation of a two-qubit system. In this part, we will look at the two-qubit teleportation protocol ($\textit{P}_1$) developed by Lee and Kim, and we will make use of two copies of the two-qubit thermal state, $\rho(T) \otimes \rho(T)$, described above as resource \cite {42}. Entanglement teleportation for the mixed channel of an intricate input state is similar to regular teleportation. After applying a local measurement in linear operators, the entangled input state is destructed, and its copy state arises at a distant location. We take a two-qubit in its peculiar pure state as input and process it. Without losing generality, let us assume that the input state is
\beq
|\psi_{in}\rangle = \cos({\theta\over2})|10\rangle +  \sin({\theta\over2})
|01\rangle, \quad (0\leq \theta \leq \pi).
\eeq
The density matrix associated with $|\psi_{in}\rangle $ has the following representation:
\begin{eqnarray} \label{input}
\rho _{in}  = \left( {\begin{array}{*{20}c}
   0 & 0 & 0 & 0  \\
   0 & \sin ^2({\theta\over2}) & \frac{1} {2}
 \sin(\theta) & 0  \\
   0 & \frac{1} {2}
\sin(\theta) & \cos ^2({\theta\over2}) & 0  \\
   0 & 0 & 0 & 0  \\
\end{array}} \right),
\end{eqnarray}
As a result, the initial state concurrence is
\beq
\mathcal{C}_{in}=\sin(\theta)
\eeq
A combined measurement and local unitary transformation on the input state $\rho_{in}$ yields the output state $\rho_{out}$. Consequently, the output state is as follows \cite {41}
\begin{eqnarray} \label{output1}
\rho _{out}  = \sum\limits_{n,m}^{} {p_{n m} (\sigma^{n}
\otimes \sigma^{m} )} \rho _{in} (\sigma ^{n}  \otimes \sigma^{m}),
\end{eqnarray}
which are denoted by the indexes $ n, m =0,x,y,z$  ($\sigma^0=I$), and  $\rho_{channel}$ indicate the  state of the channel that is being utilized for teleportation, $p_{n m}$ reflects the probability of successful teleportation as specified by
\beq
p_{n m}= \Tr [E^{n} \rho_{channel}] \Tr[E^{m} \rho_{channel}]
\eeq
where $\sum\limits_{n,m}^{} {p_{n m}}=1$, and the quantities $E^{n,m}$ are determined by the following:
\begin{eqnarray}
E^0&=&|\Psi^-\rangle\langle\Psi^-|\\   \nonumber
E^1&=&|\Phi^-\rangle\langle\Phi^-|\\  \nonumber
E^2&=&|\Phi^+\rangle\langle\Phi^+|\\ \nonumber
E^3&=&|\Psi^+\rangle\langle\Psi^+|
\end{eqnarray}
where the Bell states that they are
\begin{eqnarray}
|\Psi^\pm\rangle &=&{(|01\rangle \pm |10\rangle \over\sqrt{2}}\\  \nonumber
|\Phi^\pm\rangle &=&{(|00\rangle \pm |11\rangle \over\sqrt{2}}
\end{eqnarray}
The quantum channel state of the two-qubit spin system is $ \rho_{channel} = \rho_T$, which is provided in the equation \eqref{43}, and therefore one may derive $\rho_{out}$ as the output of the quantum channel
\begin{eqnarray} \label{output2}
\rho _{out}  = \frac{1}{\mathbb{Z}^2}\left(
\begin{array}{cccc}
 \omega & 0 & 0 & \chi \\
 0 & A^+ & B & 0 \\
 0 & B & A^- & 0 \\
 \chi & 0 & 0 & \omega \\
\end{array}
\right)
\end{eqnarray}
where
\begin{eqnarray} \label{output2 components}
 \omega &=&  4 a b \\ \nonumber
 \chi &=& 4 c d\sin (\theta )\\  \nonumber
 A^\pm &=&2 \left(\pm(a^2-b^2) \cos (\theta )+a^2+b^2\right) \\ \nonumber
 B &=& = 2 \left(c^2+d^2\right) \sin (\theta )
\end{eqnarray}
Following the procedure outlined above, we can evaluate the concurrence of the output state in order to establish the amount of entanglement associated with it by computing the square roots of the output state.
\beq\lb{RR}
{\color{black}R_{out}}=\rho _{out} S\rho _{out}^\ast S
\eeq
where $\rho _{out}$ has a complex conjugate in the form of $\rho _{out}^{\ast}$ and $S$ is calculated using the equation
\beq
S=\sigma^y\otimes\sigma^y
\eeq
where $\sigma^y$ indicates the Pauli matrix and the ${\color{black}R_{out}}$ matrix may be derived by a straightforward computation by
\beq\lb{rr}
{\color{black}R_{out}}=\frac{1}{\mathbb{Z}^4}\left(
\begin{array}{cccc}
 \chi ^2+\omega ^2 & 0 & 0 & 2 \chi  \omega  \\
 0 & A^- A^++B^2 & 2 A^+ B & 0 \\
 0 & 2 A^- B & A^- A^++B^2 & 0 \\
 2 \chi  \omega  & 0 & 0 & \chi ^2+\omega ^2 \\
\end{array}
\right)
\eeq
{\color{black}To comprehend the significance of $R_{out}$, considering that $\Tr R_{out}$, which ranges from zero to one, measures the degree of equality between $\rho_{out}$ and $\rho_{out}^\ast$, which indicates how closely $\rho_{out}$ approximates a mixture of generalized Bell states. Also, the eigenvalues of $R_{out}$ are invariant under local unitary transformations of the individual qubits, which allows them to be included in a formula for entanglement, which must also be invariant under such transformations. By using the equations \eqref{output2 components}  and \eqref{rr}, we can check quickly that
the square roots of the eigenvalues of the matrix $R_{out}$ are given by}
{\color{black}
\begin{eqnarray}\lb{R}
\lambda_{1,2} &=&\frac{4 \left(a b\pm c d C_{\text{in}}\right){}}{\mathbb{Z}^2} \\ \nonumber
\lambda_{3,4} &=&\frac{2 \left(\sqrt{\left(a^2-b^2\right)^2 C_{\text{in}}^2+4 a^2 b^2}\pm\left(c^2+d^2\right) C_{\text{in}}\right){}}{\mathbb{Z}^2} \\
\end{eqnarray}}
When trying to measure the degree of entanglement connected with $\rho_{out}$, we take into account the concurrence \cite{6,7}, which may be defined as follows:
\beq
\mathcal{C}_{out}=\max\left[0,2\max\left(\lambda_1,\lambda_2,\lambda_3,\lambda_4\right)-\sum_{i=1}^4\lambda_{i}\right]
\eeq
Consequently, we have obtained the concurrence expression for output state, which is implicitly reliant on three variables: the coupling constant $J$, the DM interaction's x-components, and the temperature $T$. As a result, we now have all of the components necessary to investigate the behavior of our proposed system concerning the previously specified quantities.
\subsection{Fidelity of entanglement teleportation}
{\color{black}The fidelity between $\rho_{in}$ and $\rho_{out}$ describes the quality of teleported state $\rho_{out}$. When the input state is a pure state, we can utilize the notion of fidelity as a useful indicator of a quantum channel's teleportation performance \cite{43, 48}.} The maximum fidelity of $\rho_{in}$ and $\rho_{out}$ is defined to be
\begin{eqnarray} \label{fidelity1}
\mathcal{F}(\rho _{in} ,\rho _{out} ) &=& \{ \Tr{[\sqrt {(\rho _{in}
)^{{\textstyle{1 \over 2}}} \rho _{out} (\rho _{in} )^{{\textstyle{1
\over 2}}} } ]\}} ^2 \nonumber \\ &=& {\color{black}\langle\psi_{in}|
\rho_{out}|\psi_{in}\rangle.}
\end{eqnarray}
By inserting $\rho_{in}$ and $\rho_{out}$ from the previous equation, we get
\begin{eqnarray} \label{fidelity2}
\mathcal{F}(\rho _{in} ,\rho _{out} ) = \frac{2  \left(a^2-b^2+c^2+d^2\right)C_{\text{in}}^2+4 b^2}{\mathbb{Z}^2}
\end{eqnarray}
In order to make the preceding formula more understandable, we may write that the maximum fidelity $\mathcal{F}(\rho_{in},\rho_{out})$ is dependent on the initial entanglement $\mathcal{C}_{in}$:
\begin{eqnarray} \label{fidelity3}
\mathcal{F}(\rho _{in} ,\rho _{out} ) =h_1 +h_2 \mathcal{C}_{in}^2
\end{eqnarray}
where $h_1=\frac{4 b^2}{\mathbb{Z}^2}$ and $h_2=\frac{2 \left(a^2-b^2+c^2+d^2\right)}{\mathbb{Z}^2}$. {\color{black}The coefficients $h_1$ and $h_2$ are related to the entanglement of the channel. This formula is the same as the results of ref. \cite{42}, but despite the Werner states, $h_1$ can be a positive number for Heisenberg chains. As a result, sending and receiving more entangled starting states with greater exactness across the same channel is possible. However, the argument is not constructive because when we choose settings for the channel such that $h_1>0$ and as long as $h_2$ decreases, the fidelity $\mathcal{F}(\rho _{in},\rho _{out})$ decreases to less than $2/3$. This imply that entanglement teleportation of a mixed state is less efficient than classical communication. Consequently, the larger-entangled channel must offer the same fidelity as a smaller-entangled channel.}

After establishing the concurrence of output state and fidelity, we will present the overall performance of the proposed model by devoting the following section to a numerical assessment of the $\mathcal{C}_{out}$ and $\mathcal{F}$, as detailed below. Following that, we will offer a few plots and continue to discuss them.

\section{Numerical results}
This section will quantitatively investigate several properties of entanglement teleportation in a two-qubit Heisenberg XXX chain with an x-component of DM  interaction. Indeed, we will investigate the output concurrence and fidelity as a function of the channel parameters {\color{black}such as temperature $ T$,  constant coupling exchange for the spin interaction $ J$ and x-component of the DM interaction $D_x$ and concurrence of the input state $\mathcal{C}_{in}$.} For simplicity, the values $k_B=\hbar=1$ will be assumed.

\begin{figure}[!h]
  \centering
  \includegraphics[width=6.8in]{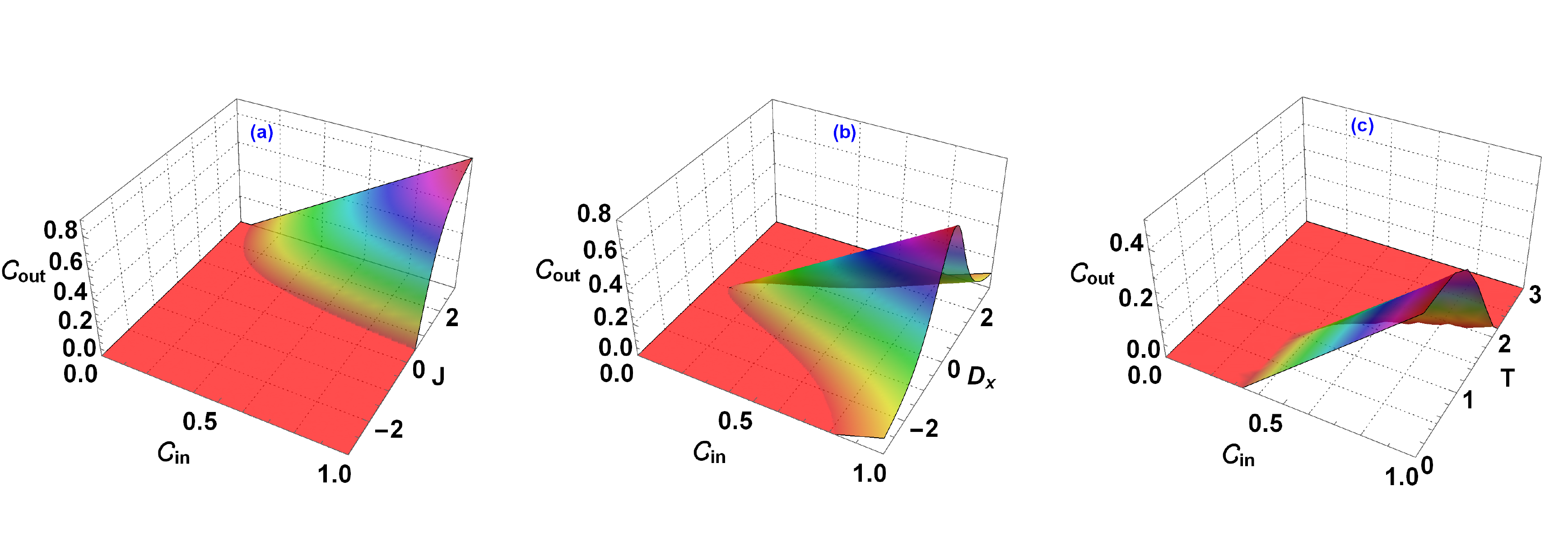}
  \caption{(Color online) Entanglement of output state $\mathcal{C}_{out}$ vs. the channel's parameters and $\mathcal{C}_{in}$. (a) $\mathcal{C}_{out}$ vs. $\mathcal{C}_{in}$ and $J$ where $D_x=T=1$, (b) $\mathcal{C}_{out}$ vs. $\mathcal{C}_{in}$ and $D_x$ where $J=T=1$, (c) $\mathcal{C}_{out}$  vs. $\mathcal{C}_{in}$ and $T$ for $J=D_x=1$.}\label{fig1}
\end{figure}
Plot \ref{fig1} displays the behavior of $\mathcal{C}_{out}$ for the channel parameters and $\mathcal{C}_{in}$. In Fig. \ref{fig1}(a), we plotted $\mathcal{C}_{out}$ versus $\mathcal{C}_{in}$, and $J$ for $D_x=T=1$. Under the constraint, $J > 0$, replicate state $\mathcal{C}_{out}$ entanglement rises linearly as $\mathcal{C}_{in}$ increases. $J$ designates the degree of this rise. However,  when $J < 0$, the concurrence $\mathcal{C}_{out}$ is zero for small values of $J$.  Fig. \ref{fig1}(b) indicates that $\mathcal{C}_{out}$ is symmetrical for the value $D_x = 0$ for any value of $\mathcal{C}_{in}$. Still, on the other hand, $\mathcal{C}_{out}$ is greatest when the spin-orbit coupling through DM interaction is zero for $\mathcal{C}_{in} = 1$, which generates the more entangled states of the system. Otherwise, when $|D_x|$ starts to rise, $\mathcal{C}_{out}$  falls, and the impact of $\mathcal{C}_{in}$ stays minimal and vice versa. However, for high values of the spin-orbit coupling, $\mathcal{C} _{out}$ goes towards zero; therefore, we can conclude in this instance that the impact of the spin-orbit coupling through DM interaction is more dominating than the concurrence of the input state. Fig. \ref{fig1}(c) demonstrates that for $ T=0$ and $ \mathcal{C}_ {in} = 1 $, the concurrence of the output state is most significant, which renders the states of the system more entangled, so when there is a rise in the temperature or deterioration of $ \mathcal{C}_{in} $ we notify a decrease of $ \mathcal{C}_ {out} $. Furthermore, when $ \mathcal{C}_ {in}\rightarrow0 $ for whichever value of $ T $ or $ T \rightarrow \infty $ and any $ \mathcal{C}_ {in} $, $ \mathcal{C}_ {out} $ tending towards zero, i.e., the output states, in this case, become separable and not entangled. To summarize,  the separability of the output states necessitates a high-temperature domain or a significant spin-orbit coupling via DM interaction or a ferromagnetic chain while obeying the abovementioned characteristics, which implies that the channel becomes disentangled, and teleportation protocol becomes impractical. But even for an antiferromagnetic chain, with weak spin-orbit coupling through DM interaction or low temperature, the system output states get more entanglement. Consequently, the channel becomes entangled, and entanglement teleportation is achievable in this case.

\begin{figure}[!h]
  \centering
  \includegraphics[width=6.8in]{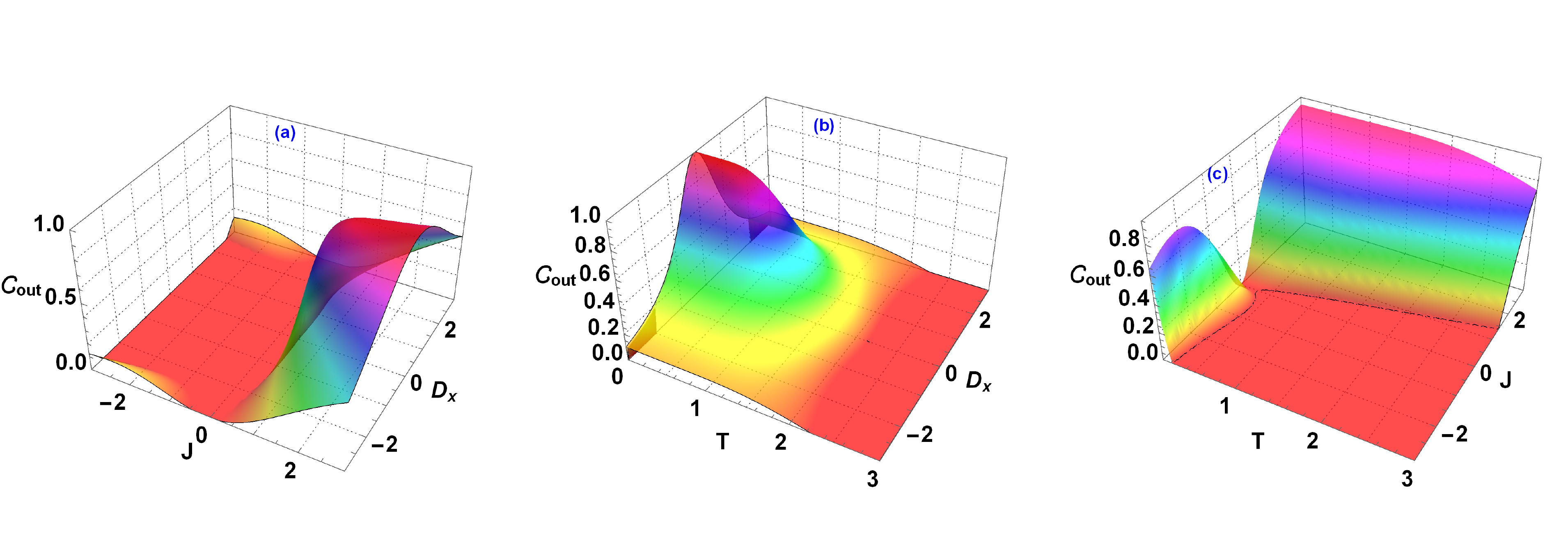}
  \caption{(Color online)(a) Entanglement of output state $\mathcal{C}_{out}$ vs. the channel's parameters and $\mathcal{C}_{in}$. $\mathcal{C}_{out}$ vs. $J$ and $D_{x}$ where $\mathcal{C}_{in}=T=1$, (b) $\mathcal{C}_{out}$ vs. $T$ and $D_{x}$ where $\mathcal{C}_{in}=J=1$, (c) $\mathcal{C}_{out}$  vs. $J$ and $T$ for $\mathcal{C}_{in}=D_x=1$.}\label{fig2}
\end{figure}
Fig. \ref{fig2} displays the behavior of $\mathcal{C} _{out}$  in terms of the channel parameters. In Fig. \ref{fig2}(a) we plot the concurrence of the output state as a function of $ J $ and $ D_ x $ for the values $\mathcal{C}_ {in}=T=1$, which is noted that for $ J<0 $, $ \mathcal{C}_ {out} $  stays zero, even though for large values of $ |D _x| $, but in the other hand for $ J>0 $, the concurrence $ \mathcal{C}_ {out} $ grows as $ J $ increases and diminishes when $ D_x $ raises. The Fig. \ref{fig2}(b), the concurrence of the output state is plotted as a function of $ T $ and $ D_x $, for $\mathcal{C} _{in}=J=1$. In these conditions, we find that $ \mathcal{C}_ {out} $ stays zero of the large values of $ D_ x $ or $ T $, which illustrates the separability of the output states in this region. However, for small values of $ D_x $ and low temperatures, $ \mathcal{C}_ {out} $ becomes maximal, and the system's output states become entangled. In Fig \ref{fig2}(c), we displayed the concurrence of the output state as a function of $T$ and $J$ by setting $\mathcal{C}_ {in}=D_ x=1$, so for the values $ J \rightarrow +\infty $ and at low temperature, $ \mathcal{C}_ {out} $ reaches maximal, and the states of the system become entangled. We have seen that $ \mathcal{C}_{out} $ stays zero for large values of $ T $ and $ J <0 $ and that the system gets increasingly entangled at low temperatures and as $ |J| $ grows. Then we may deduce that the DM interaction and temperature control the output state's system. Ferromagnetic chains and high temperatures make the system less entangled. As a result, the channel gets disentangled, and entanglement teleportation is infeasible.
\begin{figure}[!h]
  \centering
  \includegraphics[width=6.8in]{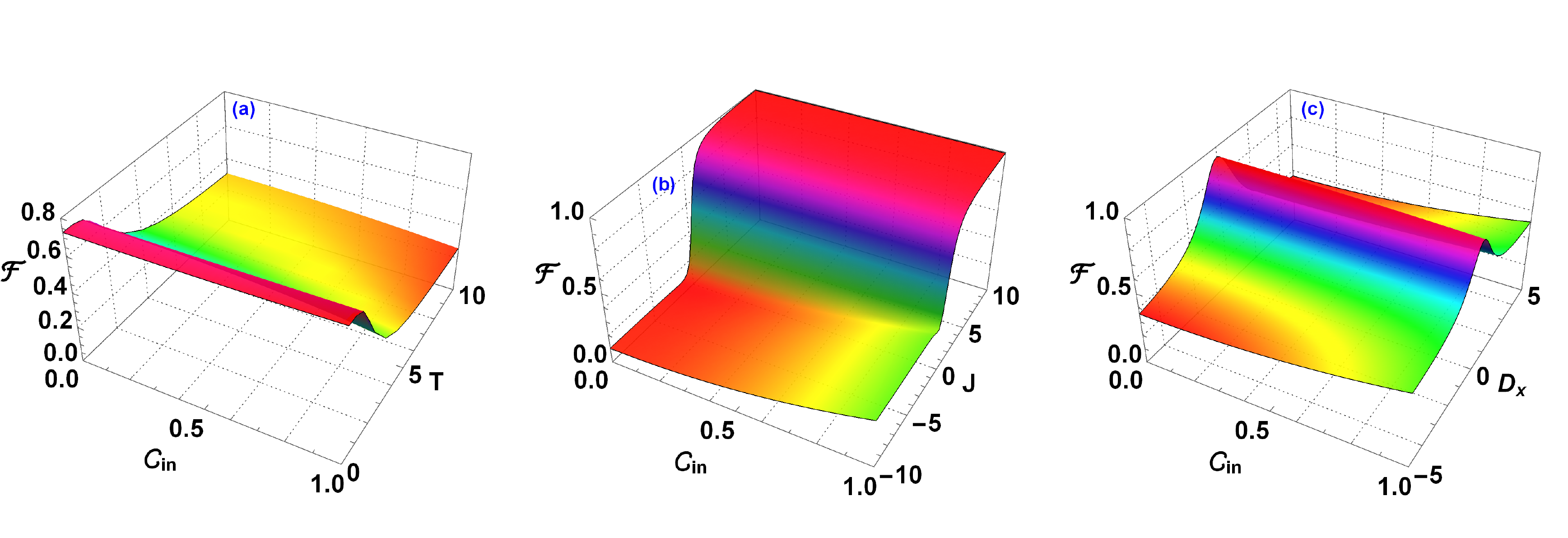}
  \caption{(Color online) The fidelity $\mathcal{F}$ vs. the channel's parameters and $\mathcal{C}_{in}$. (a) $\mathcal{F}$ vs. $\mathcal{C}_{in}$ and $T$ where $J=D_{x}=1$, (b) $\mathcal{F}$ vs. $\mathcal{C}_{in}$ and $J$ where $T=D_x=1$, (c) $\mathcal{F}$  vs. $\mathcal{C}_{in}$ and $D_x$ for $T=J=1$. }\label{fig3}
\end{figure}

Graph \ref{fig3} shows how $\mathcal{F}$ behaves concerning the channel parameters and $\mathcal{C}_ {in}$. In Fig. \ref{fig3}(a), we plotted fidelity vs. $\mathcal{C}_ {in}$ and $T$ for fixed values of $J = D_x = 1$; it is obvious that $\mathcal{F}$ stays practically constant for certain temperatures, which declines as $T$ increases. Moreover, even if the concurrence of the input state changes, the fidelity is optimum for $T = 0$. Physically, for the value $T = 0$, the fidelity $\mathcal{F}$ becomes more than $0.6$, thus making the channel suitable for performing the teleportation protocol.
In figure \ref{fig3}(b), the fidelity $\mathcal{F}$ is shown as a function of $\mathcal{C}_{in}$ and $J$ for fixed values of $T = D_x = 1$.   Particularly for high values of $J > 0$, the fidelity tends towards $1$, implying that the system is more entangled and the teleportation protocol becomes practical. Furthermore, in the case where $J < 0$, the fidelity $\mathcal{F}$ stays around $0\leq F\leq0.2$, making the system states more separable. As a result, the teleportation protocol for the proposed system is completely impractical.
Fig. \ref{fig3}(c) shows the fidelity $\mathcal{F}$ as a function of $\mathcal{C}_{in}$ and $D_ x$ for fixed values of $T = J = 1$. The fidelity reaches a maximum of $D_x = 0$, where the teleportation protocol becomes feasible. When $D_x$ is increased, the fidelity decreases to a  small amount, and the impact of $\mathcal{C}_{in}$ is insignificant.

\begin{figure}[!h]
  \centering
  \includegraphics[width=6.8in]{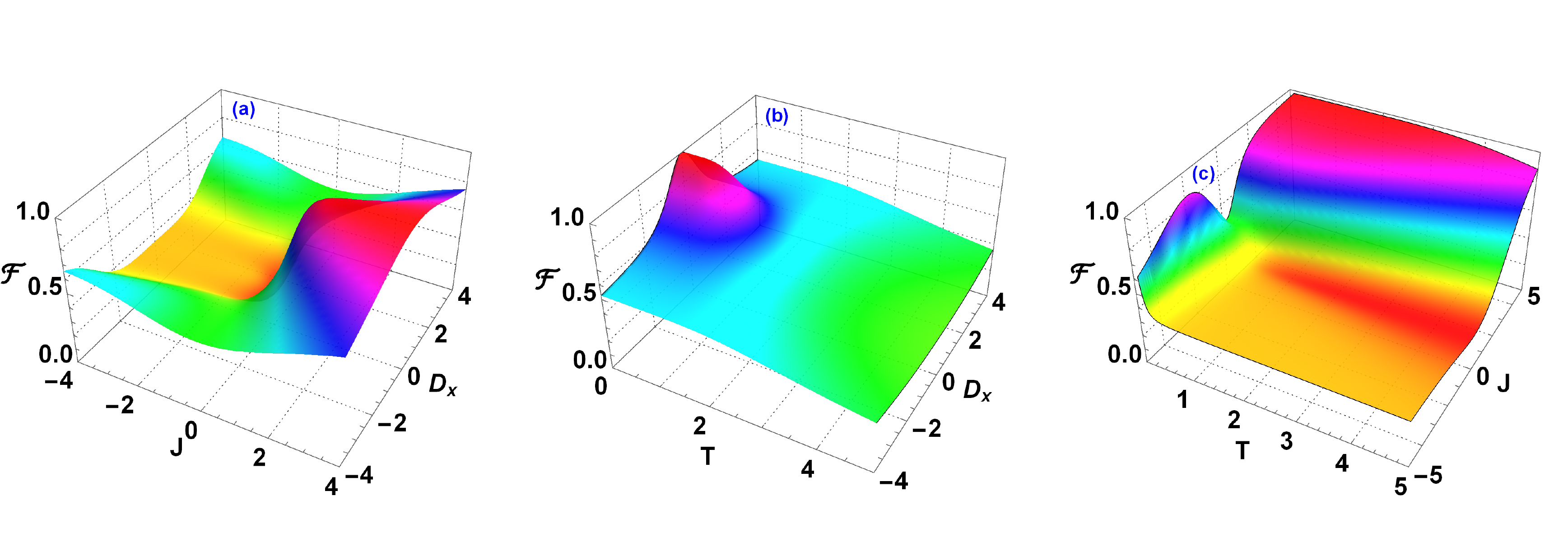}
  \caption{The fidelity $\mathcal{F}$ vs. the channel's parameters and $\mathcal{C}_{in}$. (a) $\mathcal{F}$ vs. $J$ and $D_{x}$ where $\mathcal{C}_{in}=T=1$, (b) $\mathcal{F}$ vs. $T$ and $D_{x}$ where $\mathcal{C}_{in}=J=1$, (c) $\mathcal{F}$  vs. $T$ and $J$ for $\mathcal{C}_{in}=D_{x}=1$.}\label{fig4}
\end{figure}
In Fig. \ref{fig4}, we display the fidelity $\mathcal{F}$ in terms of the channel parameters. Fig. \ref{fig4}(a) shows the fidelity in terms of $J$ and $D_x$ for fixed values of $\mathcal{C}_{in}=T=1$. In plot 4(a), the fidelity $\mathcal{F}$ always tends towards $1$ for positive $J$ values, but when $J$ turns negative, $\mathcal{F}$ takes a small value on the order of $0.2$. Moreover, for $D_x$ farther from zero, there are two observations: the first is that for $J > 0$, the fidelity decreases with rising $|Dx|$, and the second is that for $J < 0$, the fidelity grows concurrently with growing $|Dx|$.
Concerning the \ref{fig4}(b) and \ref{fig4}(c) supplement graphs \ref{fig2}(b) and \ref{fig2}(c) with almost identical findings. In conclusion, the graphs illustrated for fidelity confirm all the previous remarks, and there is conformity with all that was found for the concurrence of the output state.
\section{Summary {\color{black} and perspectives}}
{\color{black}Concurrence, a measure of entanglement of the output state and teleportation protocol fidelity, has been investigated in a two-qubit Heisenberg XXX chain with x-components of DM interaction. First, the Hamiltonian model have been presented, and the eigenstates of entanglement have been discovered using mathematical calculations. Then, the thermal state at a finite temperature is explicitly derived in terms of the parameters channel. Afterward, we established the concurrence of the output state and fidelity expressions in terms of the spin's coupling constant $J$, the x-components $D_x$ of DM interaction, the temperature $T$, the numerical behavior of the concurrence measured entanglements, and the fidelity of the teleportation protocol of our model have been investigated. Additionally, we have concluded from our research that the temperature $T$, the x-component of the DM interactions $D_x$, and the spin's coupling constant $J$ may all play an impact on influencing the entanglement output states and teleportation protocol quality to more excellent or insufficient. Furthermore, it is reasonable to conclude from these findings that the separability of the output states requires a high-temperature domain, strong spin-orbit coupling via DM interaction, or a ferromagnetic chain, which implies that the output states become nonentangled, making the teleportation protocol unfeasible and impossible in this situation. However, even in the presence of an antiferromagnetic chain, weak spin-orbit coupling through DM interaction, or low temperature, the system output states become more entangled. As a result, the channel gets entanglement, so the teleportation protocol is possible and achievable.}

{\color{black}Still some interesting questions to be addressed. Can we use the studied system to investigate the teleportation dynamic behaviors and describe the basic features of the quantum entanglement at a finite time \cite{49,50,51}? A related question arose, what about the long-range interaction as type of the Calogero–Moser interaction \cite{52}? These issues and associated questions are under consideration.}
\section*{Acknowledgment}


Thank you to Mohamed Monkad, director of the Laboratory for Physics of Condensed Matter (LPMC) at Chouaib Doukkali University's Faculty of Sciences, for his invaluable support.


\end{document}